\documentclass[journal]{IEEEtran}
\pdfoutput=1
\usepackage{ifpdf}
\ifCLASSINFOpdf
   \usepackage[pdftex]{graphicx}
   \DeclareGraphicsExtensions{.pdf,.jpeg,.png}
\else
   \usepackage[dvips]{graphicx}
   \DeclareGraphicsExtensions{.eps}
\fi
\usepackage{subfigure}
\usepackage[cmex10]{amsmath}
\usepackage{amsfonts,amssymb}
\usepackage{algorithmic}
\usepackage{array}
\usepackage{mdwmath}
\usepackage{mdwtab}
\usepackage{eqparbox}
\def\Eps{{\cal E}}

\def\CC{{\mathbb C}}

\def\ZZ{{\mathbb Z}}

\def\conv{{\bf \text{conv}\,}}

\newtheorem{theorem}{Theorem}[section]

\begin{document}

\title{Sparsity Enhanced Decision Feedback Equalization}

\author{Jovana Ilic,~\IEEEmembership{Student Member,~IEEE,}
        and~Thomas~Strohmer\thanks{Both authors were supported by NSF
project DMS 0811169 and T. Strohmer was supported  DARPA/ONR under Award No. N66001-11-1-4090.}}

\maketitle

\begin{abstract}
For single-carrier systems with frequency domain equalization, decision
feedback equalization (DFE) performs better than linear equalization and
has much lower computational complexity than sequence maximum
likelihood detection. The main challenge in DFE is the feedback symbol
selection rule. In this paper, we give a theoretical framework for a
simple, sparsity based thresholding algorithm.  We feed back multiple
symbols in each iteration, so the algorithm converges fast and has a low
computational cost. We show how the initial solution can be obtained via
convex relaxation instead of linear equalization, and illustrate the impact that the choice of the initial solution has on the bit error rate performance of our algorithm. The algorithm is applicable in several existing wireless communication systems (SC-FDMA, MC-CDMA, MIMO-OFDM). Numerical results illustrate significant performance improvement in terms of bit error rate compared to the MMSE solution.
\end{abstract}


%

\section{Introduction}
\label{s:intro}
In broadband, high data-rate, wireless communication systems, the effect of multipath propagation can be severe. While orthogonal frequency division multiplexing (OFDM) successfully deals with multipath, it is a multicarrier modulation that suffers from a large peak to average power ratio (PAPR). On the other hand, a more traditional single carrier modulation with time domain equalization approach is unattractive, due to the high complexity of the receiver and required signal processing time. When single carrier modulation is used in combination with frequency domain equalization, one attempts to approach the performance and complexity of OFDM, while maintaining a lower PAPR compared to OFDM~\cite{FABE02}.
\par
Single carrier frequency division multiple access (SC-FDMA), is a single carrier technique that has lately received much attention as an alternative to orthogonal frequency division multiple access for 4G technology. SC-FDMA has been adopted for uplink transmission technique in both 3GPP Long Term Evolution (LTE) and LTE Advanced standards~\cite{MG08}. Since most of the cost in communication terminals comes from the power amplifier, a lower PAPR can significantly reduce the cost of mobile units.  This results in a more power efficient and less complex mobile terminals. Since the orthogonal frequency division multiple access (OFDMA) is used in the downlink, both the burdens of complex frequency domain equalizer needed for the SC-FDMA and accommodating large PAPR in OFDMA rest upon the base station.  
\par
Frequency domain equalization includes frequency domain linear equalization, decision feedback equalization and turbo equalization~\cite{vazi}. For frequency selective channels, decision feedback equalization (DFE) gives much better performance than linear equalization and has a lower complexity and computational cost than optimum equalizers and turbo equalizers. The basic idea behind the DFE is to subtract (feed back) correctly equalized symbols in order to reduce the interference for the currently equalized symbols. If the wrong symbols are fed back, the interference will be further increased, so choosing which symbols are correct and should be fed back is a crucial step for any decision feedback algorithm.  Existing DFE algorithms are mostly based on finding the minimum mean square error solution (MMSE) solution of the system, and then forming some metric (such as covariance matrix, or mean square error matrix), associated with that solution. The element of the solution that corresponds to the minimum of that metric is assumed to be the one that is most likely correct, and it is fed back. The equalizer is usually implemented using a frequency domain feed-forward and time domain feed-back filter, such as in~\cite{BT02} and~\cite{HNA08}. Vertical Bell Labs Layered Space Time (V-Blast),~\cite{G98}~\cite{KP05}, has been proposed as receiver architecture for MIMO systems and can be viewed as a generalized decision feedback equalizer~\cite{GC01}. The drawback is that only one symbol is fed back in each iteration, so the complexity is linear in the block length. Even if multiple symbols are fed back, there is no general or systematic rule on how many symbols should be fed back, the number is fixed in each iteration. In this paper we address these issues with an adaptive thresholding rule for feedback symbol selection. Motivated by recent work in sparse recovery and compressive sensing~\cite{D06}, our algorithm gives a theoretical framework, based on sparsity, for multiple symbol feedback selection. Our algorithm converges in very few iterations and its performance substantially improves upon MMSE equalization. We note here that a similar concept, successive interference cancellation, exists in multiple access schemes, where users cause interference for each other. This is especially a challenge in cases, such as code division multiple access (CDMA) when there is no strict time or frequency orthogonality between different users~\cite{Verdu98},~\cite{RL00}.

\par
The rest of the paper is organized as follows. In section~\ref{s:problem} we give the problem statement. In section~\ref{s:sicthresh} we will present two ways of obtaining an initial solution for our algorithm and make the connection between sparsity of the error signal and the optimal thresholding rule for the DFE. Furthermore, we will introduce an adaptive thresholding algorithm.  Section~\ref{s:sim} is devoted to numerical results.  Finally, in section~\ref{s:con} we will give our concluding remarks and discussion of open problems.

\section{Problem Statement}
\label{s:problem}

\subsection{SC-FDMA}

While the decision feedback algorithm presented in this paper can be applied to several different technologies, such as MC-CDMA, MIMO OFDM, in this paper we focus on SC-FDMA. We will describe the SC-FDMA system model, and then explain how this model can be extended to other systems.
\begin{figure*}[!t]
\centering
\includegraphics[scale=.6]{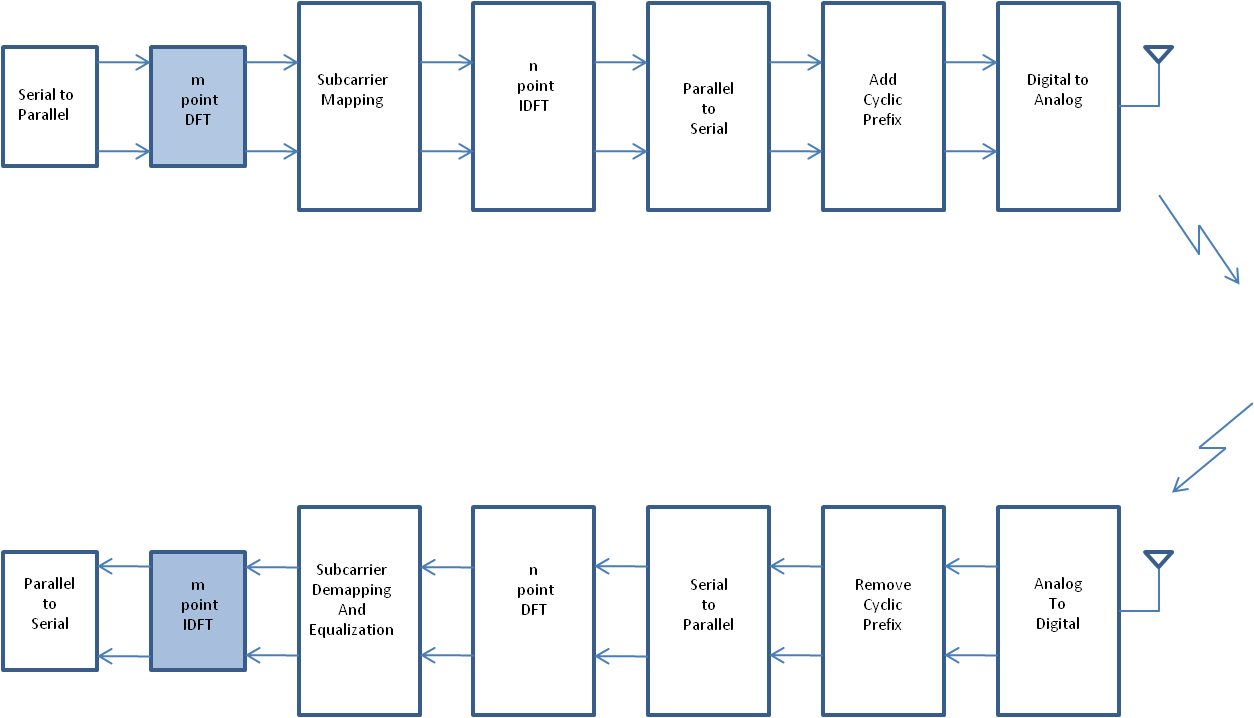}
\caption{Transmitter and Receiver Model for SC-FDMA}
\label{fig:ltemodel}
\end{figure*}   
\par
Figure~\ref{fig:ltemodel} depicts the high level model of an SC-FDMA receiver and transmitter. $m$ modulated source symbols are converted to frequency domain. The frequency domain symbols are then mapped onto $m$ out of $n$ ($m<n$) possible orthogonal subcarriers. Subcarriers can be mapped in two ways: localized mapping, where each user is assigned a set of $m$ consecutive subcarriers, and distributed mapping, where subcarriers assigned to the user are equally spaced across the entire channel bandwidth. After converting the symbols back to the time domain using an $n$-point IDFT and inserting the cyclic prefix, the SC-FDMA time domain symbol is transmitted through the channel. At the receiver all the steps are reversed. The  crucial difference between the SC-FDMA and OFDMA comes from the additional DFT block before subcarrier mapping (shaded in the figure). The DFT block "spreads" the modulated source symbols, so that each subcarrier in frequency domain contains information about all the source symbols. While this has an advantage of multipath diversity, it also destroys the decoupling of the source symbols, since we no longer have one-to-one mapping between the source symbols and subcarriers. The result is that, unlike in OFDM, simple, one-tap equalization combined with symbol-by-symbol detection is not equivalent to maximum likelihood detection (MLD). In fact, the complexity of MLD for SC-FDMA grows exponentially with the block size, $m$, making it unsuitable for practical purposes. Sphere decoding can be successfully implemented with lower complexity than MLD, however, for large block sizes, $m$, the complexity is still too high.    
\par 
It is convenient to consider a matrix formulation of an SC-FDMA system. In particular, for one user, the received vector, $Y\in \CC^m$ in time domain, (see e.g.\ equation~(11) of~\cite{HNA08}) is given by
\begin{equation}
\label{scfdma}
Y=F^{-1}(FH'F^{-1})Fx+\omega,
\end{equation}
where $F$ is an $m\times m$ DFT matrix, $H'\in \CC^m$ is a circulant channel matrix, $x\in \CC^m$ is a vector of modulated source symbols, and $w\in \CC^m$ additive white Gaussian noise (AWGN) . Since we are interested in frequency domain equalization, from~\eqref{scfdma} we can get the following
\begin{equation}
\label{scfdmafreq}
y=HFx+\omega,
\end{equation}
where $y\in \CC^m$ is a received vector for one user in frequency domain and $H \in \CC^{m\times m}$ is the diagonalized channel matrix. We assume that the channel is Rayleigh fading, and that the rows of $H$ are normalized. Defining $A=HF$, our system becomes
\begin{equation}
y=Ax+\omega.
\label{system}
\end{equation}
\par
From~\eqref{system} it is easy to see that by substituting matrix $F$
in~\eqref{scfdmafreq} with any unitary ``spreading" matrix $U$, such as a
Hadamard, Haar or random Gaussian matrix, we get a more general model. The
choice of $U$ depends on the particular system being modeled. We also note
that in this paper we assume that the receiver knows both the channel
matrix $H$ and the spreading matrix $U$. While we assumed for convenience
that $A$ is a square matrix, we emphasize that all results in our paper 
can be easily extended to the case of tall matrices $A$.

\par
Ideally, we would like to find the maximum likelihood (ML) solution
of~\eqref{system}, given by\footnote{The 2-norm of vector $a$ of length $n$ 
is denoted by $\|a\|=\sqrt{\sum_{i=1}^n |a_i|^2}$.}
\begin{equation}
x_{\text{ML}}=\underset{x\in \mathbb S^m}{\text{argmin}}\|y-Ax\|^2,
\label{ML}
\end{equation}
where $\mathbb S^m$ is the space of all vectors of length $m$ whose
elements are picked from a given constellation $\mathbb S$ (e.g., for BPSK we
have $\mathbb S=\{-1,+1\}$). As mentioned above, the ML solution is optimal,  but
the complexity of solving~\eqref{ML} grows exponentially with $m$, and
therefore it cannot be used for practical purposes even for small $m$.
While sphere decoding reduces the computational complexity
of ML considerably, it is still too costly for moderate or large $m$.
\par
In the literature, the terms equalization and detection are often (mistakenly) 
used interchangeably, but in our case it is really important to distinguish
between the two. Equalization refers to operations done on the
observation vector $y$ in order to obtain the estimate of the transmitted
vector (such as minimum mean square error equalization, or least squares
equalization). However, at this stage, the estimate still contains the
"soft" information, and not the actual symbols from the used constellation.
The mapping of the estimate into the symbols of the used constellation
(such as BPSK, or QPSK) is detection. The point of equalization is to allow
for a simple coefficient-by-coefficient detection of the equalized vector
instead of the computationally so expensive sequence detection done
in~\eqref{ML} (for ML there is of course no need for equalization, as
we immediately obtain the detected solution).
In this paper, we feed back the detected symbols, and not the soft information, so from here on, when we talk about obtaining and feeding back the initial solution, we are referring to the \textbf{detected} symbols. 

\subsection{Decision Feedback Equalization}
To explain the idea behind the decision feedback equalization, let us assume that we want to equalize the $l^{th}$ symbol in vector $x$.  We can rewrite $y$ as
$$
y=A^lx_l+ \sum_{i \in L}A^ix_i+\omega,
$$
where $L=\{i\in \ZZ \quad |\quad 0\leq i\leq n-1,\quad i\neq l \}$ and $A^l$ denotes the $l^{th}$ column of matrix $A$. The first term in the last equation is simply the symbol we want to equalize, $x_l$, scaled by the channel. The summation term, $I=\sum_{i \in L}A^ix_i$, at least as far as equalization of $x_l$ is concerned, is viewed as interference. The hope is that if we have previously correctly equalized and detected some of the $x_i$ $i\in P$, where $P\subseteq L$, we can use that knowledge to reconstruct $I_P=\sum_{i \in P}A^ix_i$ and subtract it from $y$. In this way, we are subtracting the contributions of interference from our observation. Basically, for the purpose of equalization of  $x_l$, the interference is reduced, which gives us a better chance of recovering $x_l$ correctly. In the subsequent iterations, we will have a reduced system, since we will omit the columns of $A$ that correspond to the index set of correctly equalized symbols in the previous iteration. So our system for all iterations $k>0$ will be overdetermined, which increases our likelihood of recovering correctly the remaining symbols.   
\par 
While this concept sounds very nice in theory, in practice we face a very difficult question: how do we know which symbols are equalized correctly and should be fed back? Unfortunately, there is no way to ensure that we are feeding back the correct symbols. It is even more unfortunate that if we feed back the wrong symbols, we further increase the interference and cause error propagation. Obviously, the performance of any DFE algorithm is determined by the selection rule of the feedback symbols. The other question that arises is how many symbols should we feed back in each iteration. While feeding back one symbol at a time, as is done in V-BLAST~\cite{G98}, may seem like the safest option, the computational time that it requires for larger block sizes, $m$, might  be unacceptable for some applications. Also, in a good signal to noise ratio (SNR) situation, the majority of the symbols would most likely be correct, so feeding back one symbol at a time would be a waste of resources. Hence there is a tradeoff: from the performance point of view, we would rather  feed back fewer symbols, that are guaranteed to be correct, while from a computational point of view we want to feed back as many symbols as possible in each iteration, in order to have fewer iterations. 
\par
\begin{figure}[ht]
\centering{
\subfigure[Absolute value of the true error signal in the first iteration, $|e|$]{
\includegraphics[scale=.5]{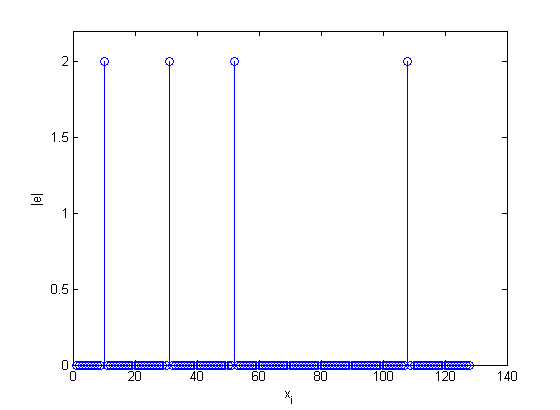}
\label{fig:subfig1}
}
\subfigure[Absolute value of the estimated error signal in the first iteration, $|\hat{e}|$]{
\includegraphics[scale=.5]{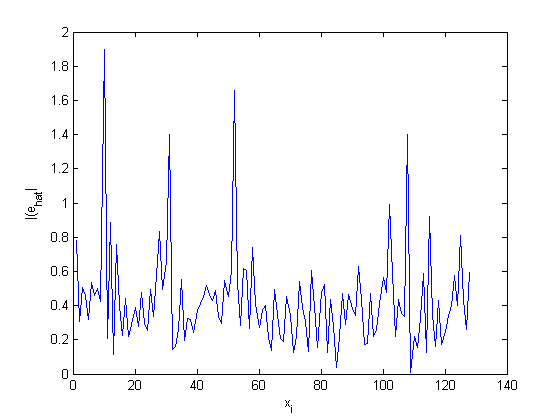}
\label{fig:subfig2}
}
\label{fig:jovana}

\caption{Comparison of the true and estimated error signals}}
\end{figure}

Let us assume for the moment that $x$ is known at the receiver. Then we would be able to compute the {\em error signal} given by
\begin{equation}
e=x-\hat{x},
\label{error}
\end{equation}
where $\hat x$ is the estimate of x obtained at the receiver after
equalization and detection. Note that for each $\hat x_i$, $i=0,...,m-1$ that matches $x_i$, the corresponding entry in vector $e_i$ would be $0$. So, assuming that we did a decent job of estimating $x$, then $e$ is a sparse vector, where the locations of the non-zero entries of $e$ correspond to the locations of errors we made in our estimate of $x$. One realization of $e$ is shown in Figure~\ref{fig:subfig1}.  We can immediately see that knowing this error vector would be ideal for our DFE selection rule: if we knew the locations of errors, we would simply not feed back the symbols that correspond to them, while we could safely feed back all symbols whose entries correspond to the zero entries of $e$. 

\par
Unfortunately, a true solution for $x$ is not known at the receiver, so we cannot construct the error signal $e$ given by~\eqref{error}. We can try to obtain an estimate $\hat{e}$ of $e$, and use this information for our feedback selection rule. One such estimate is shown in Figure~\ref{fig:subfig2}. We can see that the largest peaks in Figure~\ref{fig:subfig2} correspond to the locations of errors in Figure~\ref{fig:subfig1}. However, there are a lot of small peaks that come from the noise, and our goal is to come up with a threshold rule that will be able to distinguish the "true" peaks in the estimated error signal from the noise. Also, as we reduce the interference in the subsequent iterations the error signal will look differently, which means that the chosen threshold should adapt appropriately.  
\par 
From our previous discussion we can see that in order to design an efficient decision feedback equalization algorithm that utilizes iterative adaptive thresholding of the error signal, we need to provide answers to the following crucial questions:
\begin{enumerate}
	\item How do we find the initial solution, $\hat{x}$?
	\item How do we obtain the error estimate $\hat{e}$?
	\item How do we design a threshold that will separate true peaks from the noise, and adapt to the error signal in each iteration?
\end{enumerate}

\section{Successive Interference Cancellation with Adaptive Thresholding}
\label{s:sicthresh}

\subsection{Initial Solution via Linear Equalization}
\label{ss:insol}
In a decision feedback algorithm, in each iteration, we first must obtain
an initial solution that will be used to determine which symbols are
correctly equalized and should be fed back. Obviously, a solution closer to
the actual transmitted vector will give more accurate information for our
decision feedback rule, so obtaining a good estimate of $x$ in each
iteration obviously has an impact on the performance of our algorithm.
\par
 The simplest way to obtain $\hat{x}$ is using zero forcing (ZF)
$$
x_{\text{ZF}}=A^\ast(AA^\ast)^{-1}y,
$$
or an MMSE solution
$$
x_{\text{MMSE}}=A^\ast(AA^\ast+\sigma^2I)^{-1}y.
$$
For instance for MMSE, $\hat{x}$ is now obtained from $x_{\text{MMSE}}$
by projecting each coefficient of $x_{\text{MMSE}}$ onto $\mathbb S$.

Unfortunately large noise enhancement severely degrades the performance of ZF. MMSE offers better performance than ZF, but the ISI is still present~\cite{HNA08}. 

\subsection{Initial Solution via Convex Relaxation}
\label{ss:convex}

From a computational viewpoint the problem with the optimization 
problem~\eqref{ML} is that we need to find the minimum over 
a non-convex set, the symbol space $\mathbb S^m$. A natural idea is
then to consider a convex relaxation of~\eqref{ML} by replacing
$\mathbb S$ by its convex hull $\conv \mathbb S$ (for a definition of a convex hull see~\cite{BV04}). 
Thus instead of~\eqref{ML} we are concerned with
\begin{equation}
x_{\text{}}=\underset{x\in \conv \mathbb S^m}{\text{argmin}}\|y-Ax\|^2.
\label{infopt}
\end{equation}
Clearly, $\conv \mathbb S^m = (\conv \mathbb S)^m$. For instance for QPSK
$\conv \mathbb S = \{x\in \CC: \max\{|\Re \{x\}|, |\Im\{ x\}|\} \le 1\}$.
Thus in that case~\eqref{infopt} can be expressed as
\begin{equation}
\label{QPSK}
\min \|Ax-y\|_2\qquad \text{s.t}\qquad 
\|\Re \{x\}\|_{\infty}<1,\quad \|\Im \{x\}\|_{\infty}<1.\footnotemark
\end{equation}
\footnotetext{The infinity-norm of vector $a$ of length $n$ is given by
$\|a\|_{\infty}=\max\{|a_1|,...,|a_n|\}$.}
Some theoretical results for the noise-free, underdetermined setting
and the special case $\mathbb S = \{\pm 1\}$ can be found in~\cite{MR09,DT10}.
However, in our case the issue is not underdeterminedness, but
noise. Therefore the results in the aforementioned papers have little
bearing on our situation.

\par  
We note here that while the solution obtained via~\eqref{infopt} leads to
a better performance than MMSE (as we will show in section~\ref{s:sim}) the
computational cost of solving~\eqref{QPSK} is higher. Nevertheless, due to
recent progress in convex optimization (partly driven by the thriving area of
compressive sensing) we have now a number of fast algorithms for the
solution of problems like~\eqref{infopt}.
\par
\textit{Remark:} Because of the noise, the solution we obtain by solving~\eqref{infopt}
or~\eqref{QPSK} will not necessarily be from a finite alphabet of our constellation. So in order to obtain our
$\hat{x}$ we still have to  perform symbol by symbol detection step  as discussed in the previous section. The same is true for $x_\text{ZF}$ or $x_\text{MMSE}$.

\subsection{Error Signal And Adaptive Thresholding}
\label{ss:error}
In the area of compressed sensing greedy algorithms have been successfully
used in finding the sparsest solution for large, underdetermined systems. 
While the recovery of our error signal does not fall into the category of a 
large underdetermined system with a sparse solution, our approach is
inspired by the Stagewise Orthogonal Matching Pursuit (StOMP), an iterative
thresholding algorithm for finding  sparse solutions of~\cite{D06}. We use a similar idea for determining
which symbols in our current solution are correct and should be fed back 
in order to reduce the interference for the next iteration.
\par
Let us assume that in the $k^{th}$ iteration we have obtained $\hat{x}_k$. 
Then we can form the corresponding residual, $r_k$, as
\begin{equation}
\label{residual}
r_k=y_k-A_k\hat{x}_k,
\end{equation} 
where $A_k$ denotes the matrix that is obtained from matrix $A$ by leaving
out the columns that correspond to the index set of correctly equalized
symbols in each previous iteration (the number of rows of $A_k$ is still $m$, 
but the number of columns gets smaller in each iteration).
Then the estimate of $e$ in the $k$-th iteration is given by
\begin{equation}
\hat{e}_k=A_k^\ast r_k.
\label{eest}
\end{equation}
The key observation is that the vector $\hat{e}$ can be viewed as a sparse,
spiky signal embedded in noise, and therefore we can represent it as
\begin{equation}
\hat{e}_k=e_k+z_k,
\label{ervec}
\end{equation}
where $z_k$ is the noise term in the k-th iteration. We will later show
that under certain conditions $z$ is approximately additive white Gaussian 
noise (AWGN).
\par
Now that we were able to obtain an estimate of $e$ we need to come up with
a threshold which will help us determine which entries in $\hat{e}$ are
small enough to be considered just noise (no error was made for that index) 
and thus should be fed back. 

\par
It is a well known result, that the maximum of a random Gaussian sequence,
$c \in \mathbb C^m$, $c_k \sim {\cal CN}(0,\sigma^2)$, is bounded 
by~\cite{llr83}
\begin{equation}
\max(|c|)<\sqrt{2\sigma^2 \log m},\qquad k=0,...m-1
\label{gaus}
\end{equation}
with high probability. So if we had an unknown ``spiky'' function
embedded in AWGN,~\eqref{gaus} would be a natural choice for the threshold
that would distinguish between the spikes and the noise: we could assume
with very high probability that everything that is below~\eqref{gaus} is
indeed just noise and not a ``true'' spike. In~\cite{ABDJ00}, the authors use~\eqref{gaus} to obtain an optimal threshold rule for recovering a sparse signal embedded in AWGN noise that adapts to the level of sparsity. They modify~\eqref{gaus} by exploiting the number of spikes (level of sparsity) of the function that they are thresholding. In particular, their proposed threshold is given by
\begin{equation}
t_{\beta}=\sigma_m\sqrt{2(1-\beta) \log m},\qquad 0<\beta <1,
\label{tbeta}
\end{equation}
where $\rho=m^{\beta}$ is the level of sparsity, and $\sigma_m$ the
variance of the noise term. Via a simple calculation~\eqref{tbeta} can 
be expressed as
\begin{equation}
t_{\rho}=\sigma_m \sqrt{2 \log m/\rho},
\label{tn0}
\end{equation}
which is more convenient for our purposes.
The threshold depends on $\log m/\rho$, rather than just $\log m$, and the
penalty factor of $\log m/\rho$ accounts for the number of spikes that we
are expecting. So the more spikes we have (the less sparse the signal is), the lower the threshold gets. 
 Clearly in case that the signal has only one spike, $\rho=1$, equation~\eqref{tn0} is reduced to~\eqref{gaus}. 
\par
We emphasize here that our objective is different from the one in~\cite{ABDJ00} or~\cite{D06}: we are not interested in recovering the amplitudes of non-zero elements (spikes) of the error signal as it is the case in the compressed sensing applications. We are only interested in the positions that are zero, or very close to zero since those are the entries that we need to feed back to reduce the interference. In other words, we are only interested in locations of entries that are \textbf{below} the threshold. Furthermore, we point out that in our case, if we "miss" some zero locations in a given iteration, we do not face a performance penalty, it just means that we might have more iterations. However, if we feed back a location that is actually a spike, we increase the interference and cause error propagation. In that sense, our problem is not symmetrical,   so for our purposes, it is better to feed back fewer entries, (which corresponds to choosing a lower threshold), than to feed back the wrong entries. Obviously, the "safest" threshold rule would be to find the error estimate $\hat{e}$, and feed back only the smallest entry of $|\hat{e}|$, but then the number of iterations needed would be equal to the block length $m$. We will show in section~\ref{s:sim} that while feeding back one symbol per iteration does have a superior BER performance compared to our adaptive thresholding rule, the computational times are very high.
\par
In our case the threshold in the $k^{th}$ iteration becomes
\begin{equation}
t_k=\sqrt{2\log (m_k/\rho_k)}\sqrt{\Eps[\|z_k\|^2]}.
\label{thresh}
\end{equation}

From~\eqref{thresh} we can see that we still need to obtain the level of
sparsity, $\rho$, as well as the variance of the noise term $z_k$. The
level of sparsity is determined by the number of errors we make in our
solution. This number will be different in every iteration, so our
threshold has to adapt appropriately. We obviously cannot know the number of errors, $\rho$, that occurred in our current solution, but we need to know at least approximately the level of sparsity of the actual error vector $e$. We can obtain this estimate in the $k^{th}$ iteration as
\begin{equation}
\rho_k=\|r_k\|^2/s_{\min}^2,
\label{k}
\end{equation}
where $s_{\min}$ is the minimum distance among symbols for the used
constellation. Note that the number of unknowns decreases from one iteration 
to the next, hence the length, $m_k$ of $\hat{e}_k$ will also change in every 
iteration.  
\par 
The validity of using~\eqref{thresh} as an
optimal threshold is based on the assumption that the noise $z$
in~\eqref{ervec} is AWGN. The following theorem will show that this is
indeed the case (asymptotically) at least in the first iteration, and
therefore, using~\eqref{thresh} is justified. 

\begin{theorem}
\label{th1}
Let $z_0$ be defined as $z_0=\hat{e}_0-e_0$ where $\hat{e}^{(0)}$ and
$e^{(0)}$ are defined in~\eqref{eest} and~\eqref{error}, and $e_0$ has zero
mean. Let matrix $A$ from~\eqref{system} be a square matrix ($m=n$). Then
the entries of $z_0$, $z_{i,0}$, $i=0,...,n-1$  are asymptotically i.i.d.
normally distributed with zero mean and variance of $\Eps
[\|e\|^2]/m+\sigma^2$
\end{theorem}

\begin{IEEEproof}
For clarity of presentation, throughout this proof we will omit the iteration 
index, $0$, but we emphasize that the proof applies only to the first ($k=0$) 
iteration. 

We can write $z$ in the following way:
\begin{eqnarray}
z &=& \hat{e}-e \nonumber \\
&=&A^\ast r - e\nonumber \\
&=&A^\ast (y-A\hat{x})-e \nonumber \\
&=&A^\ast(Ax+\omega)-A^\ast A\hat{x}-e \nonumber \\
&=&A^\ast A(x-\hat{x})-e+A^\ast \omega \nonumber \\
&=&A^\ast Ae-e+A^\ast \omega \nonumber \\
&=&(A^\ast A-I)e+A^\ast \omega 
\end{eqnarray}

Since $F$ (or in more general case $U$) is a unitary matrix, and the rows of the channel matrix are normalized to have unit energy on average, we have that $(A^*A)_{ii}=1$ and we can write the $i$-th entry in $z$ as
\begin{equation}
z_{i}=\sum_{j \neq i}^m \langle A^i,A^j\rangle e_{j}+{A^i}^\ast \omega.
\label{zi}
\end{equation}
Using the Central Limit Theorem, both terms in ~\eqref{zi} will be normally
distributed in the limit since $H$, $U$ $e$, and $w$ are uncorrelated.  As a sum of two normally distributed variables, $z_i$ will also have a Normal distribution.  The mean of $z_i$ will be zero since both $e$ and $\omega$ have zero mean.   In order to find the variance of $z_i$, $\Eps [\|z_{i}\|^2]$ we first find the variance $\sigma_{1}^2$ of the first term in~\eqref{zi}. 
We have that:
\begin{equation}
 \langle A^{i},A^j\rangle =\sum_{k=0}^m a_{ki}^*a_{kj}
 \label{akj}
\end{equation}
where $A^i$ is the $i^{th}$ row, $A^j$ is the $j^{th}$ column and  $a_{ki}$ are entries of matrix $A$, respectively. Since each entry, $a_{ki}$ has a magnitude of $1/ \sqrt{m}$ on average, the variance of each entry is then
$$
\Eps[{a_{ki}^2}]=\frac{1}{m}
$$
Using the Central Limit Theorem again, the variance of~\eqref{akj} is then $1/m$. The variance of $e_j$ is $\Eps [\|e\|^2]/m$ by definition so we have that $\sigma_{1}^2$ is given by:
\begin{equation}
\sigma_{1}^2=(m-1)\frac{1}{m}\frac{\Eps [\|e\|^2]}{m}=\frac{m-1}{m^2}\Eps [\|e\|^2]\approx \frac{\Eps [\|e\|^2]}{m}
\label{sigma1}
\end{equation}

The second term in~\eqref{zi} is simply a sum of Gaussian random variables, so it remains Gaussian with zero-mean and variance $\sigma^2$.
 
So finally, we have that the variance of $z_i$ is given by:
$$
\Eps [\|z_{i}\|^2]=\frac{\Eps [\|e\|^2]}{m}+\sigma^2.
$$
\end{IEEEproof}

We have shown that the variance of the noise term in~\eqref{ervec} is
\begin{equation}
\Eps [\|z_{i}\|^2]=\frac{\Eps [\|e\|^2]}{m}+\sigma^2.
\label{variance}
\end{equation}

If there are errors in the estimated solution $\hat{x}$, we can assume,
especially for higher SNR values, that $\sigma$ is much smaller than
the term that comes from the interference in~\eqref{variance}, hence 
$$
\Eps [\|z_{i}\|^2] \approx \frac{\Eps [\|e\|^2]}{m}.
$$
Since $H$ is normalized and $U$ is unitary, there holds
$$
\Eps [\|e\|^2]\approx \Eps [\|HUe\|^2].
$$
Furthermore we have
$$ 
\Eps [\|r_{\hat{x}}\|^2]=\Eps [\|y-A\hat{x}\|^2]=\Eps[\|HUe\|^2]+\Eps [\|\omega\|^2]
$$
and thus
$$
\Eps [\|r_{\hat{x}}\|^2]=\Eps[\|HUe\|^2]+\sigma^2
$$
Using the same assumption about $\sigma$ as before, we have the following approximation
\begin{equation}
\Eps [\|e\|^2]\approx \|r_{\hat{x}}\|^2.
\end{equation}
Substituting~\eqref{k} and~\eqref{variance} into~\eqref{tn0}, we finally
obtain our threshold in the $k$-th iteration as

\begin{equation}
t_k = \sqrt{2\log (m_k/\rho_k)}\frac{\|r_k\|}{\sqrt{m_k}}
\label{threshfinal}
\end{equation}

\begin{figure}[ht]
\centering{
\subfigure[Absolute value of the true error signal in the first iteration, $|e|$]{
\includegraphics[scale=.5]{./figure2.png}
\label{fig:subfig3}
}
\subfigure[Absolute value of the estimated error signal in the first iteration, $|\hat{e}|$]{
\includegraphics[scale=.5]{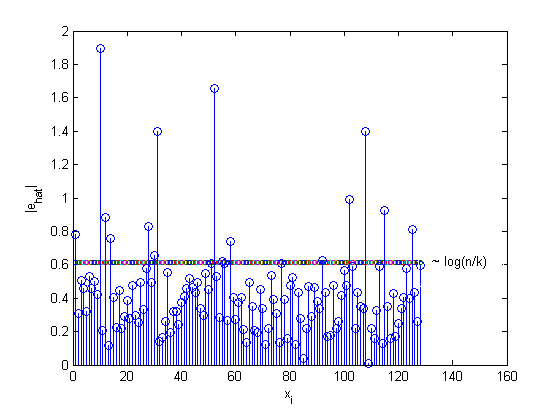}
\label{fig:subfig4}
}
\label{fig:jovana1}

\caption{Comparison of the true and estimated error signals}}
\end{figure}

In Figure~\ref{fig:subfig4} we have shown the first iteration ($k=0$) of our
thresholding algorithm for $x$ of length 128 and signal to noise ratio (SNR) of
10dB. In Figure~\ref{fig:subfig3} we can see that the actual error signal has $k=4$
non-zero values. Our estimate, obtained by~\eqref{k}, in this case is
$\hat{k}\approx 5$. The corresponding threshold, obtained by~\eqref{threshfinal}, is
$t_0\approx 0.6$. Using this threshold, in the first iteration we feed back over 100 symbols, and they are all correct. This illustrates how our algorithm gives a systematic framework of choosing as many correct symbols as possible in each iteration. 

\par

\begin{figure}[ht]
\centering
\includegraphics[scale=.6]{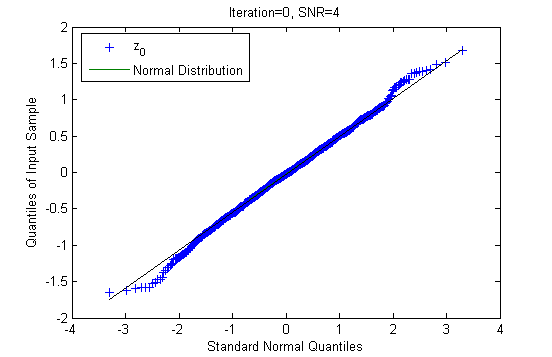}
\caption{Quantile-quantile plots of the sample quantiles of $z_0$ versus theoretical quantiles from a normal distribution}
\label{fig:qq1}
\end{figure}

\begin{figure}[ht]
\centering
\includegraphics[scale=.6]{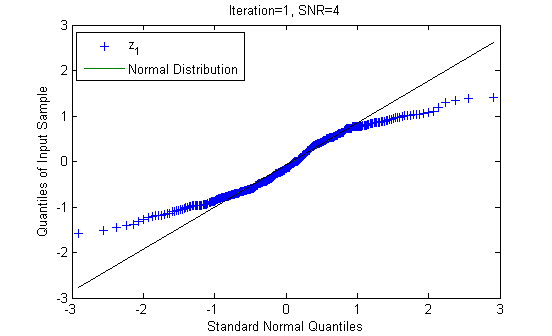}
\caption{Quantile-quantile plots of the sample quantiles of $z_1$ versus theoretical quantiles from a normal distribution}
\label{fig:qq2}
\end{figure}

\begin{figure}[ht]
\centering
\includegraphics[scale=.6]{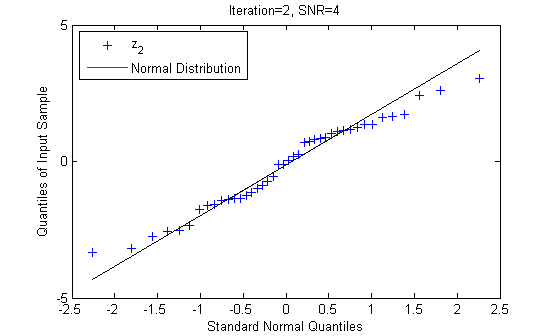}
\caption{Quantile-quantile plots of the sample quantiles of $z_21$ versus theoretical quantiles from a normal distribution}
\label{fig:qq3}
\end{figure}

We emphasize that the result in the theorem is valid \textbf{only} in the
first iteration of our thresholding algorithm. Once we start removing the
interference for the subsequent iterations, the entries in $z$ are no
longer uncorrelated. The result in Theorem 1 is significant, because it
allows us to find  the optimal threshold in the first iteration. For all
the following iterations, this threshold is no longer optimal, but our
numerical results show that the majority of the indices are fed back in the
first iteration. The chosen threshold gives satisfactory results for the
other iterations too, even though it might not be optimal. In
Figures~\ref{fig:qq1} -~\ref{fig:qq3} we have shown the quantile-quantile
plots of the sample quantiles of $z_k$ versus theoretical quantiles from a
normal distribution, for $k=0,1,2$. Figure~\ref{fig:qq1} illustrates
clearly that in the first iteration $z_0$ is very close to being
normally distributed. In Figures~\ref{fig:qq2} and~\ref{fig:qq3} we see that 
even though majority of the samples of $z_1$ coincide with the normal
distribution there are a number of entries that deviate from the normal 
distribution. The latter observation suggests that there should be room for 
improvement to our thresholding strategy. We briefly return to this issue
in our Conclusion.

\subsection{Algorithm}
\label{ss:alg}
Now that we have laid out all the necessary pieces, we are  ready to present our complete adaptive thresholding decision feedback algorithm. 
\par

From the observed vector $y$ we first obtain an initial estimate of the transmitted
vector $x$ using ZF, MMSE or convex optimization described in~\eqref{infopt}, which we  detect in order to obtain $\hat{x}_k$. 
We find the residual as
$$
r_k=y-A\hat{x}_k ,
$$
and obtain the estimate of the error signal as
$$ 
\hat{e}_k=A_k^{\ast}r_k.
$$
We calculate the threshold $t_{k}$ as 
$$
t_{k}=\sqrt{(2\log (m_k/\rho_k)}\frac{\|r_k\|}{\sqrt{m_k}}.
$$
We threshold  $|\hat{e}_k|$ and obtain the index set, $I_{c_{k}}=\{i \in \mathbb Z
|\quad |\hat{e}_{i,k}|< t_{k}\}$. The index set $I_{c_{k}}$ contains the positions of all entries in the solution $\hat{x}_k$ that are assumed to be correct. We then remove the interference caused by the ``correct" symbols:
 $$
 y_{k+1}=y_k-A_k(:,I_{c_{k}})\hat{x}_k(I_{c_{k}}).
 $$
 Here, the notation $A(:,I_c)$ denotes that all rows of $A$ are selected, but only columns that correspond to index set $I_c$ are selected. We form the matrix $A_{k+1}$ to be used in the subsequent iteration to obtain $\hat{x}_{k+1}$  by leaving out all the columns of matrix $A_k$  that correspond to index set $I_{c_{k}}$. Using $y_{k+1}$ and $A_{k+1}$ we generate the new, smaller, initial solution and repeat the process until all indices from $I=0,...m-1$ are exhausted.
 
\subsection{Error detection via $\ell_1$ Minimization}

As is the case with the initial solution, the accuracy of our error signal 
estimate can also influence the performance of our algorithm. 
Since $e$ is sparse we can attempt to approximate $e$ by
using $\ell_1$-minimization as is meanwhile common practice.

Let $\hat{x}_k$, $r_k$ be the solution and the residual in the $k$-th 
iteration, as defined in the previous subsection.
Then the estimate of $e$ in each iteration can be obtained by solving 
the following $\ell_1$-minimization problem:
\begin{equation}
\min \|\hat{e}\|_1 \quad \text{s.t.}\quad \|A\hat{e}-r_{\hat{x}} \|^2 \leq
n\sigma^2.\footnotemark
\label{l1}
\end{equation}
\footnotetext{The $\ell_1$ norm of vector $a$ of length $n$ is given by $\|a\|_1=\sum_{i=1}^n|a_i|$}
The estimate $\hat{e}$ obtained is a good approximation of the actual error
signal - at least in the high-SNR case. $\ell_1$-minimization has been 
tremendously successful in recovering sparse signals from underdetermined 
linear systems in the noise-free or high-SNR setting. However for the low-SNR 
case it is unfortunately much less effective, even though we are not dealing
with an underdetermined system. In particular, if the noise is large
enough such that $\|r_{\hat{x}}\|^2 \leq n \sigma^2$, then the optimal solution 
to~\eqref{l1} is $\hat{e}=0$, which is not useful. Since $n\sigma^2$ represents 
only the expected energy of the noise, using a more conservative choice
in~\eqref{l1}, such as  e.g.\ $\frac{n\sigma^2}{2}$, can improve the result somewhat. 
Furthermore, even though the solution obtained via~\eqref{l1} will be mostly sparse, there can be
some small, non-zero entries due to the noise, so we would still need to apply some kind of threshold before
feeding back.
However, obtaining the error estimate solution via 
$\ell_1$-minimization does not yield superior performance compared to using~\eqref{eest}, as the numerical 
simulations clearly demonstrate in the next section. Obviously, one could try to obtain the error estimate using
greedy algorithms (orthogonal matching pursuit~\cite{OMP}, subspace pursuit~\cite{SP}) used in compressed sensing as a less costly alternative to $\ell_1$ optimization, however, they did not provide any performance improvement over $\ell_1$ minimization.


\section{Simulation Results}
\label{s:sim}
In this section we present our numerical results. We consider the model as
given in~\eqref{system}. We used x with a length of 128 symbols chosen from
a QPSK constellation. The optimization toolbox CVX~\cite{cvx} has been used 
for solving both~\eqref{infopt} and the $\ell_1$ optimization problem.  

\par
\begin{figure}[!t]
\centering
\includegraphics[width=3.7in]{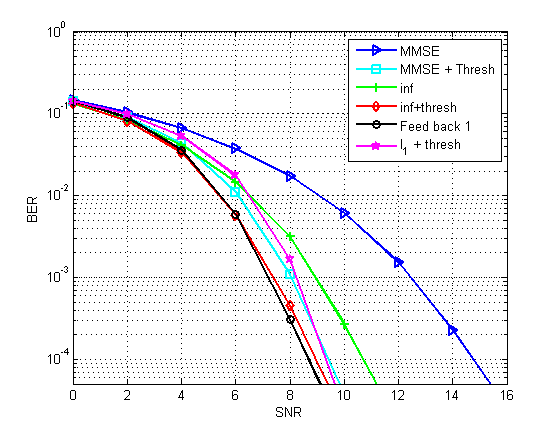}
\caption{BER performance sparsity based thresholding with different ways of obtaining initial solution and error estimate and feeding back 1 entry at a time in case $A=HU$ where $H$ is a normalized Rayleigh fading diagonal matrix, and $U$ is an DFT matrix}
\label{fig:sim1}
\end{figure}

We simulated the bit error rate (BER) performance for the following cases:
\begin{enumerate}
\item Standard linear equalizer, labeled as ``MMSE'' in the plot.
\item Solution of~\eqref{QPSK} (``inf'').
\item Our adaptive thresholding algorithm, where the initial solution is
obtained as an MMSE, and error estimate via~\eqref{eest}  (``MMSE+thresh'').
\item Our adaptive thresholding algorithm where the initial solution is obtained via optimization problem~\eqref{QPSK} and error estimate using~\eqref{eest} 
(``inf + thresh'').
\item Our adaptive thresholding algorithm, where the initial solution is obtained as an MMSE, and error estimate as using $\ell_1$ optimization~\eqref{l1} 
(``$\ell_1$ opt +thresh'').
\item Feeding back the smallest entry of $|\hat{e}|$ in each iteration 
(``Feed back'').

\end{enumerate}
Figure~\eqref{fig:sim1} depicts the results of our simulation. The first
comparison that we would like to point out is between obtaining the initial
solution using MMSE and using~\eqref{QPSK}. The performance of~\eqref{QPSK}
is significantly better - around 3.5dB at BER levels of $10^{-3}$, however,
we emphasize again that finding $\hat{x}$ using MMSE has a significantly
lower computational cost, especially when considering that an initial
solution has to be found in each iteration. We can then compare all the
thresholding scenarios. From the figure, we can see that using $\ell_1$
optimization to find the error estimate in combination with our adaptive
thresholding has an inferior performance even compared with just finding an
initial solution via~\eqref{QPSK}. The adaptive thresholding with MMSE has
a 4dB gain compared to using just MMSE at BER levels of $10^{-3}$. Adaptive
thresholding with~\eqref{QPSK} has around 0.5dB improvement compared to
using MMSE for an initial solution for BER$=10^{-3}$. Finally, we can see
that feeding back one coefficient at a time has the best performance, however, the drawback is that the number of necessary iterations is equal to the block length $m$. We note here that our adaptive thresholding algorithm usually converges within three iterations for low SNR scenarios, independently of the block size $m$. To illustrate the computational time difference between feeding back 1, our adaptive thresholding algorithm and standard MMSE, we have measured the time it took to run our simulation for 1000 QPSK symbols. Feeding back 1 took 9645 seconds, our thresholding algorithm took 321 and standard MMSE took 144 seconds. So feeding back 1 took around 30 times longer than adaptive thresholding, and around 66 times longer than MMSE.  
\par
\begin{figure}[!t]
\centering
\includegraphics[width=3.7in]{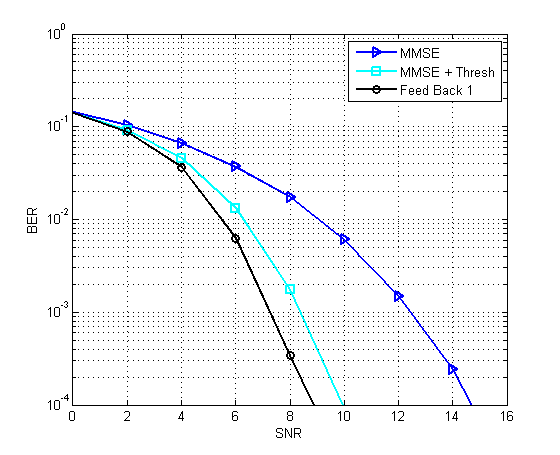}
\caption{BER performance comparison between MMSE, sparsity based thresholding and feeding back 1 entry at a time in case $A=HU$ where $H$ is a normalized Rayleigh fading diagonal matrix, and $U$ is Hadamard matrix}
\label{fig:sim2}
\end{figure}
\begin{figure}[!t]
\centering
\includegraphics[width=3.7in]{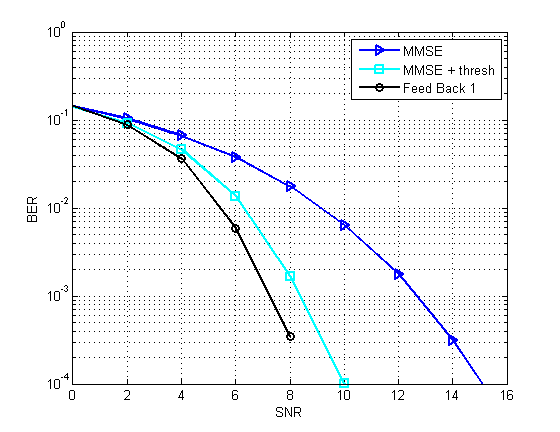}
\caption{BER performance comparison between MMSE, sparsity based thresholding and feeding back 1 entry at a time in case $A=HU$ where $H$ is a normalized Rayleigh fading diagonal matrix, and $U$ is Haar matrix}
\label{fig:sim3}
\end{figure}
\begin{figure}[!t]
\centering
\includegraphics[width=3.7in]{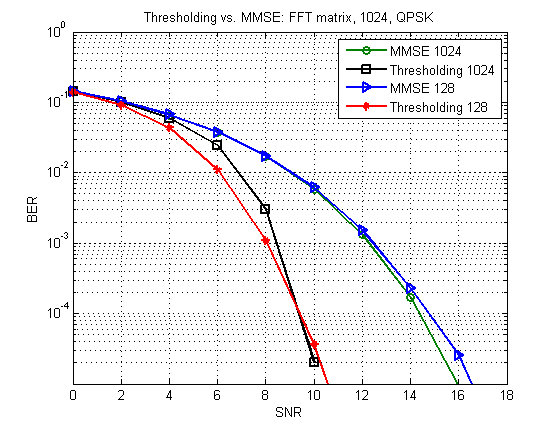}
\caption{BER performance comparison between block length of 128 and 1024 using sparsity based thresholding  in case $A=HU$ where $H$ is a normalized Rayleigh fading diagonal matrix, and $U$ is a DFT matrix}
\label{fig:sim5}
\end{figure}
\begin{figure}[!t]
\centering
\includegraphics[width=4in]{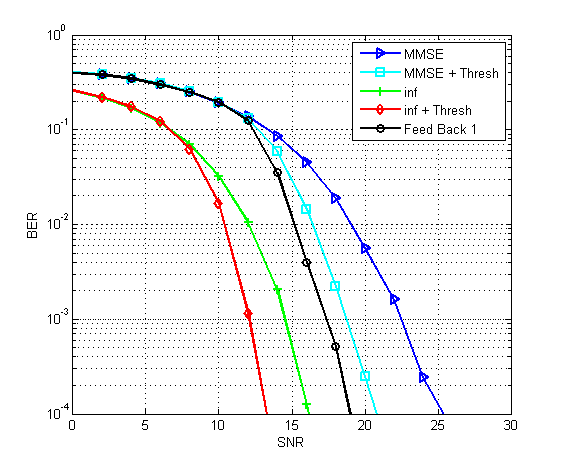}
\caption{BER performance comparison for a block length of 128 and 16-QAM modulation  in case $A=HU$ where $H$ is a normalized Rayleigh fading diagonal matrix, and $U$ is a DFT matrix}
\label{fig:sim6}
\end{figure}

From the previous discussion, we can see that in addition to the superior
performance compared to linear equalizers, our algorithm is very versatile:
depending on how we find the initial solution, the error estimate, we can
choose to sacrifice some performance in terms of BER for faster
convergence. Also, the threshold itself depends very little on the actual
system, so it can be easily adapted for different applications. In
addition, the algorithm is scalable, and can be easily be applied to larger
block sizes, with same convergence rates and performance. This is
illustrated in Figures~\ref{fig:sim2} and~\ref{fig:sim3} where we have
shown the performance of our thresholding algorithm by using Hadamard and
Haar matrices, respectively, instead of an DFT matrix in~\eqref{system}.
Our simulations show similar trends as for the DFT matrix - for
BER=$10^{-3}$ we gain about 4dB for both Hadamard and Haar matrices, by
using MMSE and adaptive thresholding, compared to just MMSE. In
Figure~\ref{fig:sim5} we have shown the performance of our algorithm with
DFT matrix, and block length of 1024. The performance improvement does not change with increasing the block size, and the algorithm still converges within 3 iterations. Finally, in Figure~\ref{fig:sim6} we show the performance of our algorithm when 16-QAM modulation is used. In this case our thresholding algorithm in combination with initial solution obtained via~\eqref{infopt} for $\text{BER}=10^{-3}$ has around 10.5dB improvement over MMSE. Unfortunately, poor MMSE performance has also significantly degraded the performance of our thresholding algorithm when the initial solution is obtained using MMSE. 
\par

\begin{figure}[!t]
\centering
\includegraphics[width=3.7in]{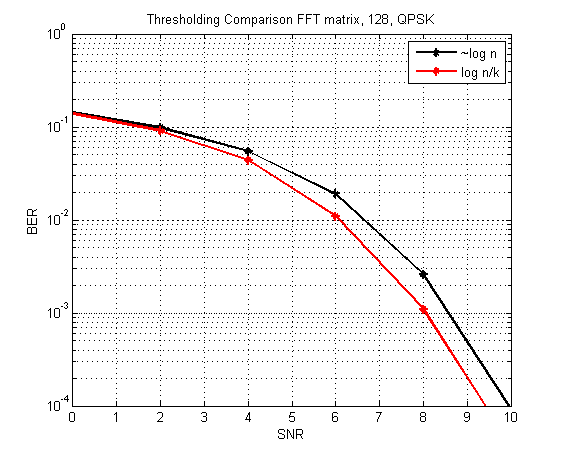}
\caption{BER performance comparison of our adaptive thresholding algorithms for threshold $\propto \log m$ and threshold $\propto \log m/\rho$}
\label{fig:sim4}
\end{figure}

In Figure~\ref{fig:sim4} we have explored the performance of threshold
given by~\eqref{gaus} and threshold given by~\eqref{threshfinal} in order
to see how much we gain by exploiting the sparsity level of the error
estimate. We can see from the figure that we gain around 0.5dB at BER=$10^{-3}$ just by introducing the penalty factor of $\log m/\rho$.

\subsection{Application To Large-System CDMA}
\label{ss:CDMA}
Here, we will briefly describe modifications needed to implement our algorithm for a large system code division multiple access and show the simulation results. We use the following system model for $K$ user system with spreading factor of $N$~\cite{GV03}:
\begin{equation}
y=SPx+\omega
\label{cdmamodel}
\end{equation}
In~\eqref{cdmamodel} $S$ represents the spreading matrix whose entries we choose from Gaussian distribution, with zero mean and unit variance. $P$ is a diagonal matrix, $P=\text{diag}(\sqrt{\Gamma_1},...,\sqrt{\Gamma_K}$, where $\Gamma_i$ denotes the signal to interference ration of user $i$. $y$, $x$ and $\omega$ are as defined in previous sections. From~\eqref{cdmamodel} we can see that if there was no interference between users, $P$ would become an identity matrix, and we would exactly get our model given in~\eqref{system}.
\par
\begin{figure}[!t]
\centering
\includegraphics[width=4in]{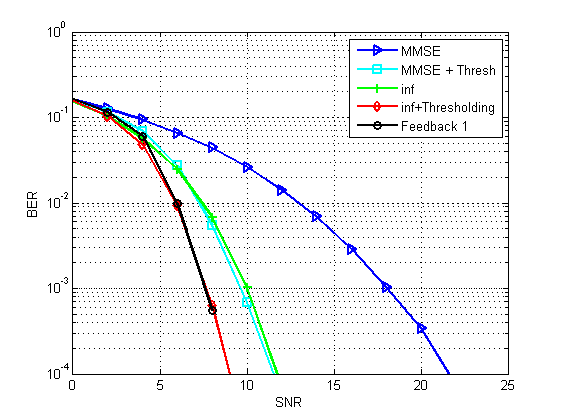}
\caption{BER performance comparison for a block length of 128 and 4-QAM modulation  in case $A=HU$ where $H$ is a normalized Rayleigh fading diagonal matrix, and $U$ is a random Gaussian matrix}
\label{fig:sim7}
\end{figure}

In Figure~\ref{fig:sim7} we have illustrated the performance of our algorithm when applied to a CDMA system given by~\eqref{cdmamodel}. We use $N=K=128$ and QPSK modulation. Matrix $S$ is a random Gaussian matrix, and it is appropriately normalized. We assume perfect power control, so we have that $\Gamma_1=...=\Gamma_i=...=\Gamma_K=\Gamma$. Figure~\ref{fig:sim7} shows that MMSE in case $U$ is random Gaussian matrix performs very poorly. That somewhat degrades the performance of our thresholding algorithm whith an MMSE initial solution, but at the BER level of $10^{-3}$ we still have an improvement of $10$dB. When an initial solution obtained via~\eqref{infopt} with our thresholding algorithm, we get the same performace as when feeding back 1 symbol at time which gives us an improvement of almost $13$dB over MMSE performance.

\section{Conclusion}
\label{s:con}
In this paper we propose a new decision feedback equalization algorithm  for SC-FDMA system. The algorithm is based on adaptive thresholding that exploits the sparsity of the estimated error signal.  We provide a theoretical framework for multiple feedback symbol selection in each iteration which leads to a very fast convergence. Our algorithm has a low computational complexity, and even though the focus of our paper is on SC-FDMA, it can easily be applied for different existing technologies such as CDMA and MIMO OFDM. We illustrated the performance of our algorithm in numerical simulations, and our algorithm shows a significant performance improvement compared to linear equalizers, while the computational time is much lower compared to feeding back one symbol at a time.
\par
While the algorithm presented in this paper offers a dramatically improved
BER performance over the linear equalizer, there is still room for
improvement, especially in the low SNR region. Recently in the area of
compressed sensing, adaptive message passing (AMP) algorithms, based on 
belief propagation, have been successfully used to improve the performance of
iterative thresholding algorithms for sparse signal recovery~\cite{D09}.
AMP can successfully account for correlations in the data, which 
is certainly of importance in our setting.
Unfortunately, we cannot simply apply the same approach, mostly because of
the step of mapping the estimated initial solution to the constellation
points. How to adapt the message passing approach to our DFE problem is
a topic of future research.

\bibliographystyle{IEEEtran}
\bibliography{IEEEabrv,JIlicTStrohmer,mathbook} 
%






\end{document}